\documentclass[prd,aps,nofootinbib,11pt]{revtex4}
\usepackage{amsmath,graphicx,color}

\begin{document}
 \title{ The $B(B_s)\to D_{(s)} P$, $D_{(s)} V$, $D_{(s)}^{*}P$ and  $D_{(s)}^{*}V$
  decays in the perturbative QCD approach}
 \author{Run-Hui Li$^{1,2}$} \author{Cai-Dian L\"{u}$^{1}$} \author{Hao Zou$^{1}$}

 \affiliation{$^{1}$Institute of High Energy Physics, P.O. Box
 918(4), Beijing 100049, China}

 \affiliation{$^{2}$Department of Physics, Shandong University,
 Jinan 250100, China}

 \date{\today}
 \begin{abstract}
Two-body non-leptonic  charmed decays $B_{(s)} \to D_{(s)}P$,
$D_{(s)}^*P$, $D_{(s)}V$ and $D_{(s)}^*V$ are analyzed in
perturbative QCD approach, where $P$ and $V$ denote the light
pseudoscalar meson and vector meson, respectively. We test the $D$
meson wave function  by a $\chi^2$ fit with experimental data of six
$B\to DP$ channels. We give the branching ratios of all the charmed
B decay channels, most of which agree with experiments amazingly
well. The predicted $B_s$ decays can be confronted with the future
experimental data. By straightforward calculations, our pQCD
approach gives the right relative strong phase of $a_2/a_1$ with
experiments. We also predict the percentage of transverse
polarizations in $B_{(s)} \to D^* V$ decay channels.
 \end{abstract}

 \maketitle

 \section{introduction}

B physics experiments provide a good  test of the standard model and
severe constraints of new physics parameters. Recent years more and
more efforts have been made to the study of B meson decays both
experimentally and theoretically.  In the near future, there will be
more and more data  in $B$  physics  thanks to the run of B
factories, Tevatron and LHCb. Theoretically, a great improvement has
been made to the study of exclusive decays of B mesons. In the
history, naive factorization \cite{bsw} is a successful method to
explain many decay branching ratios \cite{akl1}, but it failed to
explain color-suppressed processes such as $\bar B^0 \to D^0 \pi ^0$
\cite{comment on color-suppressed}. Currently, perturbative QCD
factorization approach (pQCD) \cite{Keum:2000} is one of the popular
methods to deal with the two-body non-leptonic decays of B mesons.
It explains the experiments successfully, especially for the direct
CP asymmetries \cite{direct} when the final states are two light
mesons, which inspires people to see how far it will go.

Charmed decays of $B$ ($B_s$) mesons are more complicated than the
decays with only light mesons as final states. The $B\to D$
transitions involve three scales: $M_B$, $M_D$, $\bar \Lambda$,
respectively. The factorization was proved in soft-collinear
effective theory \cite{scet} with less predictions than pQCD
approach, since they need more inputs than pQCD approach.  In $B\to$
light transition, the  light spectator quark in $B$ meson is soft,
while it is collinear in the final state meson, so that a hard gluon
is needed to connect it to the four quark operator shown in
Figure~1. The momentum square of the hard gluon connecting the
spectator quark in $B\to D$ transition is only a factor of
$(1-m_D^2/m_B^2)$ than the $B\to$ light transition, which shows pQCD
should also work well in $B\to D$ transitions. The hierarchy:
$M_B>>M_D>>\bar \Lambda$ is used in pQCD framework \cite{PQCD B to
D}.   Some separate calculations on $B\to D$ decays in pQCD approach
are carried on~\cite{Li:BtoDform,Kurimoto:2002sb}
  in the leading order  of $M_D/M_B$ and
$\Lambda /M_B$ expansions. It's found that the pQCD do work well
since the $D$ meson recoils fast.

 \begin{figure}
 \vspace{-.0cm}
 \begin{center}
 \includegraphics[scale=0.8]{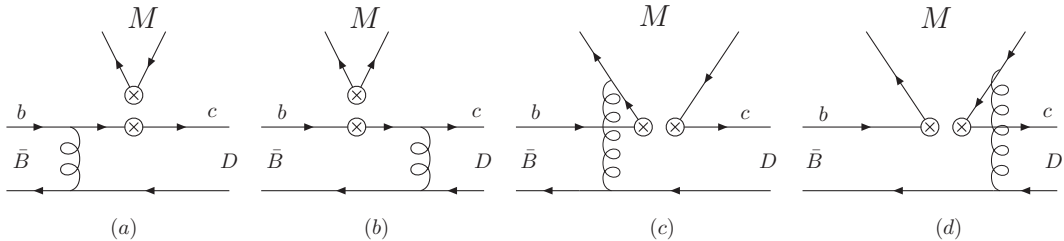}
 \caption{color allowed diagrams in pQCD approach for $B\to DP$ decays}
 \label{fig:fig_ext}
 \end{center}
 \end{figure}

In this paper, we calculate all the processes of a $B_{(s)}$ meson
decays to a $D^{(*)}_{(s)}$ meson and a light pseudoscalar meson or
vector meson. Only tree diagrams contribute to these processes
involving only one kind of CKM matrix elements. So there's no direct
CP asymmetry in these decays. The light cone distribution amplitudes
(LCDA) of mesons are necessary inputs in the pQCD framework. The
light mesons' distribution amplitudes (DAs) have been well studied
and calculated by QCD sum rules. But there are little studies on
heavy mesons' DAs, especially for D meson. In this paper we collect
several candidate distribution amplitude models for D meson, then we
fit out the parameters using the experimental results and make a
comparison among them.

 The paper
is organized as follows: Section \ref{section:analytic_formulae}
contains the conventions and notations that we adopt, together with
all the wave functions   used in this paper. The pQCD analytic
formulae for the amplitudes are given in section
\ref{section:amplitudes}. Section \ref{section:results} contain the
numerical results  and discussions. Section \ref{section:conclusion}
is our summary.

 \section{analytic formulae}\label{section:analytic_formulae}

 For the charmed B decays we considered, only the tree operators of the standard effective weak Hamiltonian
   contribute. The Hamiltonian is given by:
 \begin{equation}
 \label{effective_hamiltanian}
 {\cal
 H}_{eff}=\frac{G_F}{\sqrt{2}}V_{cb}V^*_{uq^{\prime}}\left[C_1(\mu)O_1(\mu)+C_2(\mu)O_2(\mu)\right]\;\;\;.
 \end{equation}
 The tree operators are
 \begin{eqnarray}
 O_1=(\bar c_{\alpha}b_{\beta})_{V-A}(\bar {q^{\prime}}_{\beta}u_{\alpha})_{V-A}\;\;,
 O_2=(\bar c_{\alpha}b_{\alpha})_{V-A}(\bar {q^{\prime}}_{\beta}u_{\beta})_{V-A}\;\;,
 \end{eqnarray}
 with $q^{\prime}=d,s$. $\alpha$ $\beta$ are the color indexes,
 and $(\bar q_1q_2)_{V-A}\equiv\bar
 q_1\gamma^{\mu}(1-\gamma^5)q_2$. The $V_{q_1q_2}$ in the
 Hamiltonian denote the CKM matrix elements.

 In the factorization formulae, the combinations of Wilson
 coefficients  usually appear. In this paper, they are defined as
 follows:
 \begin{eqnarray}
 a_1=C_2+C_1/3\;,\;a_2=C_1+C_2/3\;\;.
 \end{eqnarray}

 In this paper, all the momentum are defined with the light cone
  coordinate. Two light-like vectors $n$ and $v$ are defined, with $n^2=0$, $v^2=0$ and $n\cdot
  v=1$. $n$ can be written as $(1,0,\textbf{0}_T)$, and $v$ is
  $(0,1,\textbf{0}_T)$. The momentum of B meson, D meson and the lightest meson are denoted as
 $P_1$, $P_2$ and $P_3$ respectively. At the rest frame of B meson, the light meson moves very fast. So
  $P_3^+$ or $P_3^-$ can be treated as zero. In this paper, the
  momentum are defined as:
 \begin{eqnarray}
 P_1=\frac{M_B}{\sqrt{2}}(1,1,\textbf{0}_T)\;,\;
 P_2=\frac{M_B}{\sqrt{2}}(1,r^2,\textbf{0}_T)\;,\;
 P_3=\frac{M_B}{\sqrt{2}}(0,1-r^2,\textbf{0}_T)\;.
 \end{eqnarray}
 The momentum of the light anti-quark in the B meson and D meson are
 denoted as $k_1$ and $k_2$ respectively, while $k_3$ is defined as the quark
 momentum of lightest meson. In our calculation they are taken as:
 \begin{eqnarray}
k_2&=&(x_2\frac{M_B}{\sqrt{2}},0,\textbf{k}_{2\perp})\;,\;
k_3=(0,x_3\frac{(1-r^2)M_B}{\sqrt{2}},\textbf{k}_{3\perp})\;,\;\nonumber\\
k_1&=&(x_1\frac{M_B}{\sqrt{2}},0,\textbf{k}_{1\perp})\;\mbox{for
color
suppressed contributions,}\nonumber\\
k_1&=&(0,x_1\frac{M_B}{\sqrt{2}},\textbf{k}_{1\perp})\;\mbox{for
the others.}
 \end{eqnarray}
 with $x_1$, $x_2$ and $x_3$ as the momentum fraction, and
 $\textbf{k}_{1\perp}$, $\textbf{k}_{2\perp}$ and
 $\textbf{k}_{3\perp}$ as the transverse momentum of the quark.

 \subsection{Wave functions of $B_{(s)}$ mesons}

 In pQCD calculation, the light-cone wave functions of the mesons
 are needed. The $B$ meson and $B_s$ meson have the similar structure of wave
 function, except different values of parameters characterizing a small
 SU(3) breaking effect. In general, the $B_{(s)}$ meson wave function are always
 decomposed into the following Lorentz structures:
 \begin{eqnarray}
 &&\int\frac{d^4z}{(2\pi)^4}e^{ik_1\cdot z}\langle0|\bar
 b_{\alpha}(0)d_{\beta}(z)|B_{(s)}(P_1)\rangle\nonumber\\
 &=&\frac{i}{\sqrt{2N_c}}\left\{(\not P_1+M_{B_{(s)}})\gamma_5[\phi_{B_{(s)}}(k_1)-\frac{\not n-\not v}{\sqrt{2}}
 \bar\phi_{B_{(s)}}(k_1)]\right\}_{\beta\alpha}.
 \end{eqnarray}
 There are two distribution amplitudes in the above equation.
 However, $\bar\phi_{B_{(s)}}(k_1)$ gives smaller contribution \cite{phib}. We will
 neglect it in our calculation and only keep the first Lorentz
 structure.
 \begin{eqnarray}
 \Phi_{B_{(s)}}=\frac{i}{\sqrt{2N_c}}{(\not{P_1}+M_{B_{(s)}})\gamma_5\phi_{B_{(s)}}(k_1)}.
 \end{eqnarray}
 The distribution amplitude in the b-space is:
 \begin{eqnarray}
  \phi_{B_{(s)}}(x,b)&=&N_{B_{(s)}}x^2(1-x)^2\exp\left[-\frac{1}{2}(\frac{xM_{B_{(s)}}}
  {\omega_b})^2-\frac{\omega_b^2b^2}{2}\right],
  \end{eqnarray}
  with b as the conjugate space coordinate of
  $\textbf{k}_{1\perp}$. $N_{B_{(s)}}$ is the normalization constant,
  which is determined by the normalization condition:
  \begin{eqnarray}
  \int^1_0 dx\phi_{B_{(s)}}(x,b=0)=\frac{f_{B_{(s)}}}{2\sqrt{2N_c}}.
  \end{eqnarray}
 For parameter $\omega_b$, we take the value $0.40$ GeV for $B^0_d$
 and $B^{\pm}$ mesons, and $0.50\pm0.05$ GeV for $B^0_s$ meson, characterizing the small
 SU(3) breaking effect as
   argued in \cite{B_s meson decay}.

 \subsection{Wave functions of light pseudoscalar mesons}

 The decay constant of the pseudoscalar mesons is defined by:
 \begin{eqnarray}
 \langle 0|\bar
 q_1\gamma_{\mu}\gamma_5q_2|P(P_3)\rangle=if_PP_{\mu}.
 \end{eqnarray}
 The decay constant for $\pi$ and $K$ are
 $f_{\pi}=131$ MeV, $f_K=160$ MeV.

 The light cone distribution amplitudes (for out-going state) for light pseudoscalar mesons is:
 \begin{eqnarray}
 &&\langle P(P_3)|\bar
 q_{2\beta}(z)q_{1\alpha}(0)|0\rangle\\ \nonumber
 &=&-\frac{i}{\sqrt{2N_C}}\int_0^1dx
 e^{ixP\cdot z}\left[\gamma_5\not {P_3}\phi^A(x)+\gamma_5m_0\phi^P(x)+m_0\gamma_5(\not v\not
 n-1)\phi^T(x)\right]_{\alpha\beta}\;,
 \end{eqnarray}
where  $v$ is the light cone direction that the light pseudoscalar
meson's momentum is defined
 on. The chiral scale parameter $m_0$ is defined as $m_0=\frac{M_P^2}{m_{q_1}+m_{q_2}}$.

 The distribution amplitudes are usually expanded by the
 Gegenbauer polynomials. Their expressions are:
 \begin{eqnarray}
 \phi_P^A(x)&=&\frac{3f_P}{\sqrt{2N_c}}x(1-x)\left[1+a_1^AC_1^{3/2}(t)+a_2^AC_2^{3/2}(t)
                                               +a_4^AC_4^{3/2}(t)\right],\\
 \phi_P^p(x)&=&\frac{f_P}{2\sqrt{2N_c}}\left[1+a_2^pC_2^{1/2}(t)+a_4^pC_4^{1/2}(t)\right],\\
 \phi_P^T(x)&=&-\frac{f_P}{2\sqrt{2N_c}}\left[C_1^{1/2}(t)+a_3^TC_3^{1/2}(t)\right],
 \end{eqnarray}
 with t=2x-1.
 The coefficients of the Gegenbauer polynomials are \cite{pdas}
 \begin{eqnarray}
 a^A_{2\pi}&=& 0.44\;,\;a^A_{4\pi}=0.25\;,\;a^A_{1K}=0.17\;,\;a^A_{2K}=0.2\;,\nonumber\\
 a^p_{2\pi}&=& 0.43\;,\;a^p_{4\pi}=0.09\;,\;a^p_{2K}=0.24\;,\;a^p_{4K}=-0.11\;,\nonumber\\
 a^T_{3\pi}&=& 0.55\;,\;a^T_{3K}=0.35\;.
 \end{eqnarray}

 For $\eta$ and $\eta^\prime$, the mixing mechanism must be taken
 into consideration. We will take the method presented in ref.
 \cite{etamixing},   where the $\eta_n$ and $\eta_s$ are chosen as the basis of mixing:
\begin{equation}
   \left( \begin{array}{c}
    |\eta\rangle \\ |\eta^\prime\rangle
   \end{array} \right)
   = U(\phi)
   \left( \begin{array}{c}
    |\eta_{n}\rangle \\ |\eta_s\rangle
   \end{array} \right),
\end{equation}
with
\begin{equation}
|\eta_n\rangle=\frac{1}{\sqrt{2}}(\bar uu+\bar
dd)\;,\;|\eta_s\rangle=\bar ss\;,
\end{equation}
\begin{equation}
U(\phi)=\left( \begin{array}{cc}
    \cos\phi & ~-\sin\phi \\
    \sin\phi & \phantom{~-}\cos\phi
   \end{array} \right)\;,
\end{equation}
where the mixing angle $\phi=39.3^\circ\pm1.0^\circ$.

The assumption that the distribution amplitudes of $\eta_n$ and
$\eta_s$ is the same as the distribution amplitudes of $\pi$ is
adopted, except different decay constants and chiral parameters. The
decay constant of $\eta_n$ and $\eta_s$ is taken from
ref.\cite{etamixing}:
\begin{eqnarray}
 f_n=(1.07\pm0.02)f_{\pi}=(139.1\pm 2.6)~{\rm MeV}, \ \ \
 f_s=(1.34\pm0.06)f_\pi=(174.2 \pm 7.8)~{\rm MeV}.
 \end{eqnarray}
 And the chiral parameters are given by
 \begin{eqnarray}
  m_0^{\bar {n} n}=\frac{1}{2m_n}[m^2_\eta\cos^2\phi+
  m_{\eta'}^2\sin^2\phi-\frac{\sqrt 2f_s}{f_n}(m_{\eta'}^2-m_\eta^2)\cos\phi\sin\phi],\\
  m_0^{\bar{s} s}=\frac{1}{2m_s}[m^2_{\eta'}\cos^2\phi+
  m_{\eta}^2\sin^2\phi-\frac{f_n}{\sqrt
  2f_s}(m_{\eta'}^2-m_\eta^2)\cos\phi\sin\phi].
 \end{eqnarray}

 \subsection{Wave functions of light vector mesons}

 \begin{table}
 \caption{The decay constants of vector mesons (in MeV)}
 \begin{tabular}{cccccccc}
 \hline\hline
 \ \ \ $f_{\rho}$  &$f_{K^*}$  &$f_{\omega}$  &$f_{\phi}$  &$f_{\rho}^T$  &$f_{K^*}^T$  &$f_{\omega}^T$  &$f_{\phi}^T$\\
 \hline
 \ \ \ $209\pm2$   &$217\pm5$  &$195\pm3$     &$231\pm4$   &$165\pm9$     &$185\pm10$   &$151\pm9$       &$186\pm9$\\
 \hline
 \end{tabular}\label{vector_decay_constants}
 \end{table}

 The decay constants for the vector mesons are defined by:
 \begin{equation}
 \langle 0|\bar q_1\gamma_\mu
 q_2|V(p,\epsilon)\rangle=f_Vm_V\epsilon_\mu,\;\;\; \langle 0|\bar
 q_1\sigma_{\mu\nu}q_2|V(p,\epsilon)\rangle =if^T_V(\epsilon_\mu
 p_\nu-\epsilon_\nu p_\mu).
 \end{equation}
 The longitudinal decay constant of vector mesons can be extracted
 experimentally \cite{vector_decay_constant_extract}. And the
 transverse ones can be calculated by the QCD sum
 rule \cite{vector_decay_constant_sumrule}. We list all of them in
 Table~\ref{vector_decay_constants}.

 The distribution amplitudes up to twist-3 of vector mesons are
 \begin{eqnarray}
 \langle V(P,\epsilon^*_L)|\bar
 q_{2\beta}(z)q_{1\alpha}(0)|0\rangle&=&\frac{1}{\sqrt{2N_C}}
 \int_0^1dxe^{ixP\cdot z}\left[M_V\not\epsilon^*_L\phi_V(x)+\not\epsilon^*_L\not
 P\phi_V^t(x2)+M_V\phi_V^s(x)\right]_{\alpha\beta},\nonumber\\
 \langle V(P,\epsilon^*_T)|\bar
 q_{2\beta}(z)q_{1\alpha}(0)|0\rangle&=&\frac{1}{\sqrt{2N_C}}
 \int_0^1dxe^{ixP\cdot z}\left[M_V\not\epsilon^*_T\phi_V^v(x)+\not\epsilon^*_T\not
  P\phi_V^T(x)\right.\nonumber\\
  &&\left.+M_Vi\epsilon_{\mu\nu\rho\sigma}\gamma_5\gamma^\mu\epsilon^{*\nu}_Tn^\rho v^\sigma
  \phi_V^a(x)\right]_{\alpha\beta},
 \end{eqnarray}
 where x is the momentum fraction of the $q_2$ quark. And convention
 $\epsilon^{0123}=1$ is adopted for the Levi-Civita tensor.
 The twist-2 distribution amplitudes of vector mesons are expanded
 as:
 \begin{eqnarray}
 \phi_V(x)=\frac{3f_V}{\sqrt{2N_C}}x(1-x)\left[1+a^\parallel_1C_1^{\frac{3}{2}}(t)
                             +a^\parallel_2C_2^{\frac{3}{2}}(t)\right],\nonumber\\
 \phi_V^T(x)=\frac{3f_V}{\sqrt{2N_C}}x(1-x)\left[1+a^\perp_1C_1^{\frac{3}{2}}(t)
                             +a^\perp_2C_2^{\frac{3}{2}}(t)\right].\label{vwavef1}
 \end{eqnarray}
 We take the following values for the Gegenbauer moments \cite{vdas}:
 \begin{eqnarray}
 a_{2\rho}^\parallel=a_{2\omega}^\parallel=0.15\pm0.07\;,\;a_{1K^*}^\parallel=0.03\pm0.02\;,\;
 a_{2K^*}^\parallel=0.11\pm0.09\;,\;a_{2\phi}^\parallel=0.18\pm0.08\;,\;\nonumber\\
 a_{2\rho}^\perp=a_{2\omega}^\perp=0.14\pm0.06\;,\;a_{1K^*}^\perp=0.04\pm0.03\;,\;
 a_{2K^*}^\perp=0.10\pm0.08\;,\;a_{2\phi}^\perp=0.14\pm0.07\;.\;
 \end{eqnarray}
 For the other distribution amplitudes, we use the asymptotic form:
 \begin{eqnarray}
 &\phi_V^t(x)=\frac{3f_V^T}{2\sqrt{6}}t^2\;,\;&\phi_V^s(x)=\frac{3f_V^T}{2\sqrt{6}}(-t)\;,\nonumber\\
 &\phi_V^v(x)=\frac{3f_V}{3\sqrt{6}}(1+t^2)\;,\;&\phi_V^a(x)=\frac{3f_V}{4\sqrt{6}}(-t)\;.\;\label{vwavef2}
 \end{eqnarray}

\subsection{Wave function of $D^{(*)}$ meson}

The two-particle light-cone distribution amplitudes of $D^{(*)}$
meson, up to twist-3 accuracy, are defined by \cite{PQCD B to D}:
 \begin{eqnarray}
\int \frac{d^4w}{(2\pi)^4}e^{ik \cdot w}\langle 0|\bar{c}_\beta
(0) d_\gamma(w)|D^-(P)\rangle  & = &
-\frac{i}{\sqrt{2N_C}}[(\not{P}+M_D)\gamma_5]_{\gamma\beta}
\phi_D(x), \nonumber\\
 \int \frac{d^4w}{(2\pi)^4}e^{ik \cdot
w}\langle 0|\bar{c}_\beta (0) d_\gamma(w)|D^{*-}(P)\rangle  & = &
-\frac{i}{\sqrt{2N_C}}[(\not{P}+M_{D^*})\not{\epsilon}_L\phi_{D^*}^L(x)
+(\not{P}+M_{D^*})\not{\epsilon}_T\phi_{D^*}^T(x)]_{\gamma\beta}\nonumber\\
 \end{eqnarray}
 with
 \begin{eqnarray}
  \int_0^1 dx\phi_D(x)&=&\frac{f_D}{2\sqrt{2N_c}}\;,
  \int_0^1dx\phi_{D^*}^L(x)=\frac{f_{D^*}}{2\sqrt{2N_c}}\;,
  \int_0^1dx\phi_{D^*}^T(x)=\frac{f_{D^*}^T}{2\sqrt{2N_c}}\;,
  \end{eqnarray}
  as the normalization conditions. In the heavy quark limit we
  have
  \begin{equation}
  f_{D^*}^T-f_{D^*}\frac{m_c+m_d}{M_{D^*}}\sim
  f_{D^*}-f_{D^*}^T\frac{m_c+m_d}{M_{D^*}}\sim
  O(\bar\Lambda/M_{D^*}).
  \end{equation}
 Thus we use $f_{D^*}^T=f_{D^*}$ in the calculation.
 There are several candidate distribution amplitude models for D
 meson. We collect as below:
 \begin{eqnarray}
\phi_D^{(Gen)}(x,b) &=&
\frac{1}{2\sqrt{2N_c}}f_D6x(1-x)[1+C_D(1-2x)],\nonumber \\
\phi_D^{(MGen)}(x,b) &=& \frac{1}{2\sqrt{2N_c}}f_D6x(1-x)[1+C_D(1-2x)]\exp [\frac{-\omega^2 b^2}{2}],\nonumber\\
\phi_D^{(KLS)}(x,b) &=& \frac{1}{2\sqrt{2N_c}}f_DN_D\sqrt{x(1-x)}
\exp[-\frac{1}{2}(\frac{xM_D}{\omega})^2-\frac{\omega^2b^2}{2}],\nonumber\\
\phi_D^{(GN)}(x,b) &=& \frac{1}{2\sqrt{2N_c}}f_DN_Dx\exp[-\frac{xM_D}{\omega}]\frac{1}{1+b^2\omega^2},\nonumber\\
\phi_D^{(KKQT)}(x,b) &=& \frac{1}{2\sqrt{2N_c}}f_DN_Dx\theta(x)
\theta(\frac{2\Lambda_D}{M_D}-x)J_0(b\sqrt{x(\frac{2\Lambda_D}{M_D}-x)}),\nonumber\\
\phi_D^{(Huang)}(x,b) &=&
\frac{1}{2\sqrt{2N_c}}f_DN_Dx(1-x)\exp[-\Lambda_D\frac{(1-x)m_d^2+xm_c^2}{x(1-x)}].
 \end{eqnarray}
In the above models x is the momentum fraction of the light quark in
D meson. The first DA model $\phi_D^{(Gen)}$ was proposed in
\cite{PQCD B to D}, which is  the Gegenbauer polynomial-like form.
In order to make it $k_\perp$ dependent, an exponential term is
added to get $\phi_D^{(MGen)}$. The third candidate DA model
$\phi_D^{(KLS)}$ was proposed in \cite{third D meson wave function},
which is a Gaussian type model. The fourth one \cite{fourth D meson
wave function}, which is an exponential model, and the fifth model
\cite{fifth D meson wave function}, which is obtained by solving the
equations of motion without three-parton contributions, were first
proposed for B meson. Here we use heavy quark symmetry and modify
the parameters to make them D meson DAs. The sixth DA was proposed
in \cite{Huang D meson wavefunction}, which is derived from the BHL
prescription \cite{BHL}, with $m_d=0.35$ GeV, $m_c=1.3$  GeV. In the
above candidate DAs, only the second model has two parameters, and
only $\phi_D^{(Gen)}$ and $\phi_D^{(Huang)}$ are b independent. In
the next section we will try to fit out the best D meson wave
function parameters with the experimental results. As for $D^{*}$
meson, we just assume that $\phi_D^L = \phi_D^T = \phi_D$ according
to heavy quark symmetry.

 \section{Calculation of decay amplitudes in pQCD approach}\label{section:amplitudes}

 \subsection{Amplitudes for $B_{(s)}\to D_{(s)}P$ decays}

  \begin{figure}
 \vspace{-0.0cm}
 \begin{center}
 \includegraphics[scale=0.8]{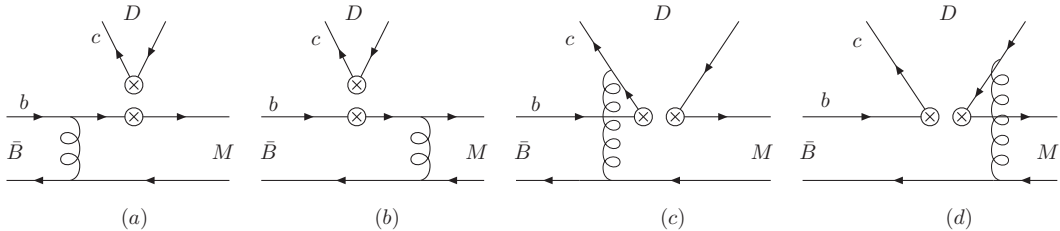}
 \caption{color suppressed diagrams  in pQCD approach for $B\to DP$ decays}
 \label{fig:fig_int}
 \end{center}
 \end{figure}

  \begin{figure}
 \vspace{-0.0cm}
 \begin{center}
 \includegraphics[scale=0.8]{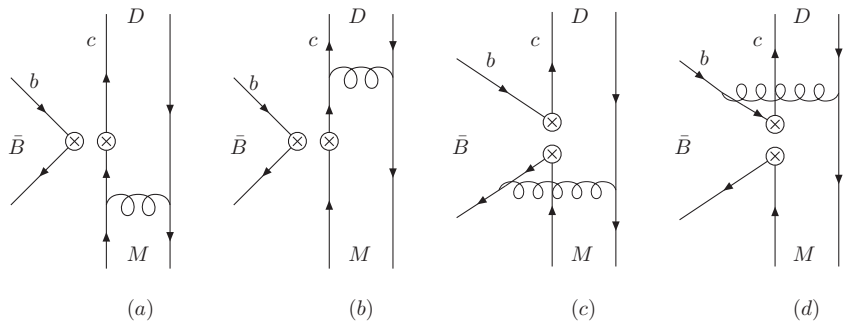}
 \caption{annihilation type diagrams  in pQCD approach for $B\to DP$ decays}
 \label{fig:fig_exc}
 \end{center}
 \end{figure}

 There are three types of diagrams that may contribute to the
 $B\to D^{(*)}M$ decays: color allowed diagrams (we mark
 this kind of contribution with the subscript $ext$) shown in
 Fig.~\ref{fig:fig_ext}, color suppressed diagrams (marked with $int$) shown
  in Fig.~\ref{fig:fig_int},
  and the annihilation type diagrams (marked with exc) shown in Fig.~\ref{fig:fig_exc}.
  And each type of diagrams contains two categories: the
  one with one meson can be factorized out
    (denoted as $\xi$) and the one with no meson can be  factorized out   (denoted as $M$).

   The first two diagrams in Fig.~\ref{fig:fig_ext}, \ref{fig:fig_int} and \ref{fig:fig_exc}
   involve only two meson wave functions, whose
 results are as following:
  \begin{eqnarray}
 \xi_{\rm ext} &=&8\pi C_Ff_P \int_{0}^{1}d x_{1}d
 x_{2}\int_{0}^{1/\Lambda} b_1d b_1 b_2d b_2
 \phi_B(x_1,b_1)\phi_{D}(x_2)
 \nonumber \\
 & &\times \left[ E_e(t_e^{(1)})h(x_1,x_2,b_1,b_2)S_t(x_2)(1+x_2+r) +
 rE_e(t_e^{(2)})h(x_2,x_1,b_2,b_1)S_t(x_1)\right]\;,
 \label{ext}\\
 \xi_{\rm int}&=&8\pi C_Ff_D \int_0^1
 dx_1dx_3\int_0^{1/\Lambda}b_1db_1b_3db_3 \phi_B(x_1,b_1)
 \nonumber \\
 & &\times \{[(2-x_3) \phi_P(x_3) - r_0 (1- 2x_3)
 (\phi_P^p(x_3)- \phi_P^T(x_3))]
 \nonumber\\
 & &\times E_i(t_i^{(1)})h(x_1,(1-x_3)(1-r^2),b_1,b_3)S_t(x_3)
 \nonumber \\
 & &\ \ + 2r_0 \phi_P^p(x_3) E_i(t_i^{(2)})
 h(1-x_3,x_1(1-r^2),b_3,b_1)S_t(x_1)\}\;,
 \label{int} \\\
 \xi_{\rm exc}&=&8\pi C_Ff_B \int_0^1
 dx_2dx_3\int_0^{1/\Lambda}b_2db_2b_3db_3 \phi_{D}(x_2)
 \nonumber \\
 & &\times \left[-x_3\phi_P(x_3)
 E_a(t_a^{(1)}) h_a(x_2,x_3(1-r^2),b_2,b_3)S_t(x_3)\right.
 \nonumber \\
 & &\ \ \  \left.+x_2\phi_P(x_3)
 E_a(t_a^{(2)}) h_a(x_3,x_2(1-r^2),b_3,b_2)S_t(x_2)\right]\;,
 \label{exc}
 \end{eqnarray}
 with the mass ratio $r_0 \equiv m_0/m_B$. $f_P$, $f_B$ and $f_D$
 are the decay constants of the light pseudoscalar meson, B meson
 and D meson respectively.
 And the factor
 evolving with the scale $t$ are given by:
 \begin{eqnarray}
 E_e(t)&=&\alpha_s(t)a_1(t)\exp[-S_B(t)-S_{D}(t)]\;,
 \nonumber \\
 E_i(t)&=&\alpha_s(t)a_2(t)\exp[-S_B(t)-S_P(t)]\;,
 \nonumber \\
 E_a(t)&=&\alpha_s(t)a_2(t)\exp[-S_{D}(t)-S_P(t)]\;.\label{Sudakov1}
 \end{eqnarray}
  We adopt the expression of Sudakov factor for D meson as suggested
  in ref.~\cite{PQCD B to D}, which is listed in
 appendix together with the expressions for $S_B(t)$, $S_P(t)$.

 The functions $h$'s in the hard part of factorization formulae,
 derived from the factorizable diagrams, are given by
 \begin{eqnarray}
 h(x_1,x_2,b_1,b_2)&=&K_{0}\left(\sqrt{x_1x_2}m_Bb_1\right)
 \nonumber \\
 & &\times \left[\theta(b_1-b_2)K_0\left(\sqrt{x_2}m_B
 b_1\right)I_0\left(\sqrt{x_2}m_Bb_2\right)\right.
 \nonumber \\
 & &\left.+\theta(b_2-b_1)K_0\left(\sqrt{x_2}m_Bb_2\right)
 I_0\left(\sqrt{x_2}m_Bb_1\right)\right]\;,
 \label{hd1}\\
 h_a(x_2,x_3,b_2,b_3)&=&\left(i\frac{\pi}{2}\right)^2
 H_0^{(1)}\left(\sqrt{x_2x_3}m_Bb_2\right)
 \nonumber \\
 & &\times\left[\theta(b_2-b_3)
 H_0^{(1)}\left(\sqrt{x_3}m_Bb_2\right)
 J_0\left(\sqrt{x_3}m_Bb_3\right)\right.
 \nonumber \\
 & &\left.+\theta(b_3-b_2)H_0^{(1)}\left(\sqrt{x_3}m_Bb_3\right)
 J_0\left(\sqrt{x_3}m_Bb_2\right)\right]\;,\label{hd2}
 \end{eqnarray}
 where $H^{(1)}(z)=J_0(z)+iY_0(z)$. The hard scales $t$ are determined by
 \begin{eqnarray}
 & &t_e^{(1)}={\rm max}(\sqrt{x_2}m_B,1/b_1,1/b_2)\;, \;\;\;
 t_e^{(2)}={\rm max}(\sqrt{x_1}m_B,1/b_1,1/b_2)\;,
 \nonumber \\
 & &t_i^{(1)}={\rm max}(\sqrt{(1-x_3)(1-r^2)}m_B,1/b_1,1/b_3)\;,
 \;\;\; t_i^{(2)}={\rm max}(\sqrt{x_1(1-r^2)}m_B,1/b_1,1/b_3)\;,
 \nonumber\\
 & &t_a^{(1)}={\rm max}(\sqrt{x_3(1-r^2)}m_B,1/b_2,1/b_3)\;,\;\;\;
 t_a^{(2)}={\rm max}(\sqrt{x_2(1-r^2)}m_B,1/b_2,1/b_3)\;.\label{hardscale1}
 \end{eqnarray}

 The formulae for the last two diagrams in Fig.~\ref{fig:fig_ext},
 Fig.~\ref{fig:fig_int} and Fig.~\ref{fig:fig_exc}   contain the
 kinematics variables of the three mesons.   Their expressions
 are:
 \begin{eqnarray}
 {\cal M}_{\rm ext}&=& 16\pi\sqrt{2N_c} C_F \int_0^1
 [dx]\int_0^{1/\Lambda} b_1 db_1 b_3 db_3
 \phi_B(x_1,b_1)\phi_{D}(x_2)\phi_P(x_3)
 \nonumber \\
 & &\times \left[x_3E_b(t_b^{(1)})h^{(1)}_b(x_i,b_i)
 -(1-x_3+x_2)E_b(t_b^{(2)})h^{(2)}_b(x_i,b_i) \right]\;,
 \label{mb}\\
 {\cal M}_{\rm int}&=& 16\pi\sqrt{2N_c} C_F \int_0^1
 [dx]\int_0^{1/\Lambda}b_1 db_1 b_2 db_2 \phi_B(x_1,b_1)\phi_D(x_2)
 \nonumber \\
 & &\times
 \left[\left((x_3-1-x_2)\phi_P(x_3)+r_0(1-x_3)(\phi_p^p(x_3)-\phi_P^T(x_3))\right)
 E_d(t_d^{(1)})h^{(1)}_d(x_i,b_i)\right.\nonumber\\
 &
 &\left.+\left[(1-x_2)\phi_P(x_3)+r_0(x_3-1)(\phi_P^p(x_3)+\phi_P^T(x_3))\right]
 E_d(t_d^{(2)})h^{(2)}_d(x_i,b_i)\right]\;,
 \label{md}\\
 {\cal M}_{\rm exc}&=& 16 \pi\sqrt{2N_c} C_F \int_0^1
 [dx]\int_0^{1/\Lambda}b_1 db_1 b_2 db_2
 \phi_B(x_1,b_1)\phi_{D}(x_2)
 \nonumber \\
 & &\times \left[x_3\phi_P(x_3) E_f(t_f^{(1)})h^{(1)}_f(x_i,b_i)
 -x_2\phi_P(x_3) E_f(t_f^{(2)})h^{(2)}_f(x_i,b_i) \right]\;,
 \label{mf}
 \end{eqnarray}
 with $[dx]\equiv dx_1dx_2dx_3$. The expressions for the evolution factors are
 \begin{eqnarray}
 E_b(t)&=&\alpha_s(t)\frac{C_1(t)}{N_C}\exp[-S(t)|_{b_2=b_1}]\;,
 \nonumber \\
 E_d(t)&=&\alpha_s(t)\frac{C_2(t)}{N_C}\exp[-S(t)|_{b_3=b_1}]\;,
 \nonumber \\
 E_f(t)&=&\alpha_s(t)\frac{C_2(t)}{N_C}\exp[-S(t)|_{b_3=b_2}]\;,\label{Sudakov2}
 \end{eqnarray}
 with the Sudakov exponent $S=S_B+S_{D}+S_P$.

 The functions $h^{(j)}$, $j=1$ and $2$, in
 these amplitudes are
 \begin{eqnarray}
 h^{(j)}_b&=&
 \left[\theta(b_1-b_3)K_0\left(Bm_B
 b_1\right)I_0\left(Bm_Bb_3\right) +\theta(b_3-b_1)K_0\left(Bm_B b_3\right)
 I_0\left(Bm_B b_1\right)\right]
 \nonumber \\
 &  & \times \left\{ \begin{array}{cc}
 K_{0}(B_{j}m_Bb_{3}) &  \mbox{for $B^2_{j} \geq 0$}  \\
 \frac{i\pi}{2} H_{0}^{(1)}(\sqrt{|B_{j}^2|}m_Bb_{3})  &
 \mbox{for $B^2_{j} \leq 0$}
  \end{array} \right\}\;,\label{hb}
 \\
 h^{(j)}_d&=&
 \left[\theta(b_1-b_2)K_0\left(Dm_B
  b_1\right)I_0\left(Dm_Bb_2\right) +\theta(b_2-b_1)K_0\left(Dm_B b_2\right)
 I_0\left(Dm_B b_1\right)\right]
 \nonumber \\
 &  & \times \left\{ \begin{array}{cc}
 K_{0}(D_{j}m_Bb_{2}) &  \mbox{for $D^2_{j} \geq 0$}  \\
 \frac{i\pi}{2} H_{0}^{(1)}(\sqrt{|D_{j}^2|}m_Bb_{2})  &
 \mbox{for $D^2_{j} \leq 0$}
 \end{array} \right\}\;,
 \label{hd}\\
 h^{(j)}_f&=& i\frac{\pi}{2}
 \left[\theta(b_1-b_2)H_0^{(1)}\left(Fm_B
 b_1\right)J_0\left(Fm_Bb_2\right) +\theta(b_2-b_1)H_0^{(1)}\left(Fm_B b_2\right)
 J_0\left(Fm_B b_1\right)\right]\;  \nonumber \\
 &  & \times \left\{ \begin{array}{cc}
 K_{0}(F_{j}m_Bb_{1}) &  \mbox{for $F^2_{j} \geq 0$}  \\
 \frac{i\pi}{2} H_{0}^{(1)}(\sqrt{|F_{j}^2|}m_Bb_{1})  &
 \mbox{for $F^2_{j} \leq 0$}
  \end{array} \right\}\;,\label{hf}
 \end{eqnarray}
 with the variables
 \begin{eqnarray}
 B^{2}&=&x_{1}x_{2}\;,
 \nonumber \\
 B_{1}^{2}&=&x_{1}x_{2}-x_2x_{3}(1-r^{2})\;,
 \nonumber \\
 B_{2}^{2}&=& x_{1}x_{2}-x_2(1-x_{3})(1-r^{2})\;,
 \nonumber \\
 D^{2}&=&x_{1}(1-x_{3})(1-r^{2})\;,
 \nonumber \\
 D_{1}^{2}&=&(x_{1}-x_{2})(1-x_{3})(1-r^{2})\;,
 \nonumber \\
 D_{2}^{2}&=&(x_{1}+x_{2})r^{2}-(1-x_{1}-x_{2})(1-x_{3})(1-r^{2})\;,
 \nonumber \\
 F^{2}&=&x_{2}x_{3}(1-r^{2})\;,
 \nonumber \\
 F_{1}^{2}&=&x_2(x_1-x_3(1-r^2))\;,
 \nonumber\\
 F_{2}^{2}&=&1-(1-x_2)(1-x_1-x_3(1-r^2))\;. \label{mis}
 \end{eqnarray}
 The scales $t^{(j)}$ are given by
 \begin{eqnarray}
 t_b^{(j)}&=&{\rm max}(Bm_B,\sqrt{|B_j^2|}m_B,1/b_1,1/b_3)\;,
 \nonumber \\
 t_d^{(j)}&=&{\rm max}(Dm_B,\sqrt{|D_j^2|}m_B,1/b_1,1/b_2)\;,
 \nonumber \\
 t_f^{(j)}&=&{\rm
 max}(Fm_B,\sqrt{|F_j^2|}m_B,1/b_1,1/b_2)\;.\label{hardscale2}
 \end{eqnarray}

 The decay amplitudes of each $B_{(s)}\to D_{(s)}P$  channels are then
\begin{eqnarray}
 A\left(B^-\to D^0\pi^-\right)&=&\frac{G_F}{\sqrt{2}}V_{cb}V^*_{ud}
        \left(\xi_{ext}+{\cal M}_{ext}+\xi_{int}+{\cal M}_{int}\right)\;,\label{B_to_Dpi-}\\
 A\left(B^-\to D^0K^-\right)&=&\frac{G_F}{\sqrt{2}}V_{cb}V^*_{us}
        \left(\xi_{ext}+{\cal M}_{ext}+\xi_{int}+{\cal M}_{int}\right)\;,\\
 A\left(\bar B^0\to D^+\pi^-\right)&=&\frac{G_F}{\sqrt{2}}V_{cb}V^*_{ud}
        \left(\xi_{ext}+{\cal M}_{ext}+\xi_{exc}+{\cal M}_{exc}\right)\;,\\
 A\left(\bar B^0\to D^+K^-\right)&=&\frac{G_F}{\sqrt{2}}V_{cb}V^*_{us}
        \left(\xi_{ext}+{\cal M}_{ext}\right)\;,\\
 A\left(\bar B^0\to D^+_sK^-\right)&=&\frac{G_F}{\sqrt{2}}V_{cb}V^*_{ud}
        \left(\xi_{exc}+{\cal M}_{exc}\right)\;,\\
 A\left(\bar B^0\to D^0\pi^0\right)&=&\frac{G_F}{\sqrt{2}}V_{cb}V^*_{ud}
        \frac{1}{\sqrt{2}}\left(-(\xi_{int}+{\cal M}_{int})+(\xi_{exc}+{\cal M}_{exc})\right)\;,\\
 A\left(\bar B^0\to D^0\bar K^0\right)&=&\frac{G_F}{\sqrt{2}}V_{cb}V^*_{us}
        \left(\xi_{int}+{\cal M}_{int}\right)\;,\\
 A\left(\bar B^0\to D^0\eta_n\right)&=&\frac{G_F}{\sqrt{2}}V_{cb}V^*_{ud}
        \frac{1}{\sqrt{2}}\left(\xi_{int}+{\cal M}_{int}+\xi_{exc}+{\cal M}_{exc}\right)\;,\label{B_to_etan}\\
 A\left(\bar B_s^0\to D^+\pi^-\right)&=&\frac{G_F}{\sqrt{2}}V_{cb}V^*_{us}
        \left(\xi_{exc}+{\cal M}_{exc}\right)\;,\\
 A\left(\bar B_s^0\to D^+_s\pi^-\right)&=&\frac{G_F}{\sqrt{2}}V_{cb}V^*_{ud}
        \left(\xi_{ext}+{\cal M}_{ext}\right)\;,\\
 A\left(\bar B_s^0\to D^+_sK^-\right)&=&\frac{G_F}{\sqrt{2}}V_{cb}V^*_{us}
        \left(\xi_{ext}+{\cal M}_{ext}+\xi_{exc}+{\cal M}_{exc}\right)\;,\\
 A\left(\bar B_s^0\to D^0\pi^0\right)&=&\frac{G_F}{\sqrt{2}}V_{cb}V^*_{us}
        \frac{1}{\sqrt{2}}\left(\xi_{exc}+{\cal M}_{exc}\right)\;,\\
 A\left(\bar B_s^0\to D^0\bar K^0\right)&=&\frac{G_F}{\sqrt{2}}V_{cb}V^*_{ud}
        \left(\xi_{int}+{\cal M}_{int}\right)\;,\\
 A\left(\bar B_s^0\to D^0\eta_n\right)&=&\frac{G_F}{\sqrt{2}}V_{cb}V^*_{us}
        \frac{1}{\sqrt{2}}\left(\xi_{exc}+{\cal M}_{exc}\right)\;,\label{Bs_to_etan}\\
 A\left(\bar B_s^0\to D^0\eta_s\right)&=&\frac{G_F}{\sqrt{2}}V_{cb}V^*_{us}
        \left(\xi_{int}+{\cal M}_{int}\right)\;.\label{Bs_to_etas}
 \end{eqnarray}
 It should be noticed that, in~(\ref{B_to_etan}),
 (\ref{Bs_to_etan}) and (\ref{Bs_to_etas}), the decay amplitudes
 are for the mixing basis of $\eta$ and $\eta^{\prime}$.   For the
 physical state
 $\eta$ and $\eta^{\prime}$, the decay amplitudes are
 \begin{eqnarray}
 A\left(\bar B^0\to D^0\eta\right)&=&A\left(\bar B^0\to D^0\eta_n\right)\cos\phi\;,\\
 A\left(\bar B^0\to D^0\eta^{\prime}\right)&=&A\left(\bar B^0\to D^0\eta_n\right)\sin\phi\;,\\
 A\left(\bar B_s^0\to D^0\eta\right)&=&A\left(\bar B_s^0\to D^0\eta_n\right)\cos\phi-A\left(\bar B_s^0\to
                                        D^0\eta_s\right)\sin\phi\;,\\
 A\left(\bar B_s^0\to D^0\eta^{\prime}\right)&=&A\left(\bar B_s^0\to D^0\eta_n\right)\sin\phi+A\left(\bar B_s^0\to
                                        D^0\eta_s\right)\cos\phi\;.\label{Bs_to_eta'}
 \end{eqnarray}

 \subsection{Amplitudes for $B_{(s)}\to D_{(s)}V$ and $B_{(s)}\to D_{(s)}^*P$ decays}

 For the channels of $B_{(s)}\to D_{(s)}V$ and $D_{(s)}^*P$, no transverse
 polarization of the vector mesons will contribute. In the leading
 powder contribution, the formulae of $B\to DV$ and $B\to D^* P$ are
 the
 same as that of $B_{(s)}\to D_{(s)}P$ decays, except some substitutions.

 For $B_{(s)} \to D_{(s)}V$, the following substitutions should be done for the formula $\xi_i$ and ${\cal M}_i$:
 \begin{eqnarray}
 \phi_P\to\phi_V\;,\; \phi_P^p\to-\phi_V^s\;,\;
 \phi_P^t\to-\phi_V^t\;,\; m_0\to m_V\;,\;
 f_P\to f_V\;.\label{replace1}
 \end{eqnarray}
 $\phi_V$, $\phi_V^s$ and $\phi_V^t$ are the light cone
 distribution amplitudes of vector mesons, which we defined
 before. And $m_V$, $f_V$ are the mass and the decay constant
 of the vector meson related.

 For $B_{(s)}\to D_{(s)}^*P$, the substitutions in the formula $\xi_i$ and ${\cal M}_i$ should be done as:
 \begin{eqnarray}
 m_D\to m_{D^*}\;,\;\;f_{D}\to f_{D^*}\;,\;\; \phi_{D}(x_2)\to
 \phi_{D^*}(x_2)\;. \label{replace2}
 \end{eqnarray}

 Making the following substitutions in equations~(\ref{B_to_Dpi-})-(\ref{Bs_to_eta'}),
  we can get the final decay
 amplitude for each $B\to D^*P$ decays:
 \begin{equation}
 D^+\;\to\;D^{*+}\;,\;D^0\;\to\;D^{*0}\;,\;D_s^+\;\to\;D_s^{*+}\;.\label{D*replace}
 \end{equation}
 And the formulae for $B\to DV$ can be obtained through the
 substitutions
 \begin{equation}
 \pi\;\to\;\rho\;,\;K\;\to\;K^*\;,\;\eta_n\;\to\;\omega\;,\;\eta_s\;\to\;\phi\label{Vreplace}
 \end{equation}
 in equations (\ref{B_to_Dpi-})-(\ref{Bs_to_etas}).

 \subsection{Amplitudes for $B_{(s)}\to D_{(s)}^*V$ decays}

 For $B\to D^*_{(s)}V$ decays, both longitudinal and transverse
 polarization can contribute. For the longitudinal polarization,
  the amplitudes can be obtained by carrying out the
 substitutions referred in~(\ref{replace1})
 and~(\ref{replace2}), when only the leading power contribution is
 taken into consideration. The transverse polarized contribution
 is suppressed by $r$ or $r_V$, $r_V\equiv m_V/m_B$. Although the
 transverse polarization will not give the leading power contribution,
  we   still list the analytic formulae  for transverse polarization
 $\xi_{\rm ext}^T$, $\xi_{\rm int}^T$ and $\xi_{\rm exc}^T$
 \begin{eqnarray}
 \xi_{\rm ext}^T &=&8\pi C_Fm_B^4f_V \int_{0}^{1}d x_{1}d
 x_{2}\int_{0}^{1/\Lambda} b_1d b_1 b_2d b_2
 \phi_B(x_1,b_1)\phi_{D}^T(x_2)
 \nonumber \\
 & &\times r_V\bigg[
 E_e(t_e^{(1)})h(x_1,x_2,b_1,b_2)S_t(x_2)\left(-\epsilon^{n\bar n\epsilon_D^{*T}\epsilon_V^{*T}}-i\epsilon_D^{*T}
 \cdot\epsilon_V^{*T}(1+2r)\right)\nonumber\\
&&
+E_e(t_e^{(2)})h(x_2,x_1,b_2,b_1)S_t(x_1)r(r+1)\left(-\epsilon^{n\bar
n\epsilon_D^{*T}\epsilon_V^{*T}}-i\epsilon_D^{*T}
 \cdot\epsilon_V^{*T}\right)\bigg]\;,
 \label{text}\\
 \xi_{\rm int}^T&=&8\pi C_Fm_B^4f_{D^*} \int_0^1
 dx_1dx_3\int_0^{1/\Lambda}b_1db_1b_3db_3 \phi_B(x_1,b_1)
 \nonumber \\
 & &\times r\bigg\{\left[\epsilon^{n\bar n\epsilon_D^{*T}\epsilon_V^{*T}}\left(-\phi_V^t(x_3)
 -r_V((x_3-3)\phi_V^a(x_3)+(x_3-1)\phi_V^v(x_3))\right)\right.\nonumber\\
 &&\left.-i\epsilon_D^{*T}
 \cdot\epsilon_V^{*T}\left(\phi_V^t(x_3)-r_V((x_3-1)\phi_V^a(x_3)-(x_3-3)\phi_V^v(x_3)))\right)\right]
 \nonumber\\
 & &\times E_i(t_i^{(1)})h(x_1,x_3(1-r^2),b_1,b_3)S_t(x_3)
 \nonumber \\
 & &\ \ + r_V\left[\epsilon^{n\bar n\epsilon_D^{*T}\epsilon_V^{*T}}(\phi_V^a(x_3)-\phi_V^v(x_3))+i\epsilon_D^{*T}
 \cdot\epsilon_V^{*T}(\phi_V^a(x_3)-\phi_V^v(x_3))\right]\nonumber\\
 &&\times E_i(t_i^{(2)})h(x_3,x_1(1-r^2),b_3,b_1)S_t(x_1)\bigg\}\;,
 \label{tint} \\\
 \xi_{\rm exc}^T&=&8\pi C_Fm_B^4f_B \int_0^1
 dx_2dx_3\int_0^{1/\Lambda}b_2db_2b_3db_3 \phi_{D}^T(x_2)
 \nonumber \\
 & &\times \left[
 E_a(t_a^{(1)}) h_a(x_2,x_3(1-r^2),b_2,b_3)S_t(x_3)\right.\nonumber\\
 &&\left.\times\big(\epsilon^{n\bar
 n\epsilon_D^{*T}\epsilon_V^{*T}}[r^2\phi_V^t(x_3)-rr_V(x_3+1)\phi_V^a(x_3)+rr_V(1-x_3)\phi_V^v(x_3)]\right.
 \nonumber\\
& &\left.+i\epsilon_D^{*T}
 \cdot\epsilon_V^{*T}[-r^2\phi_V^t(x_3)+rr_V(x_3-1)\phi_V^a(x_3)+rr_V(x_3+1)\phi_V^v(x_3)]\big)\right.
 \nonumber \\
 & &+\left.
 E_a(t_a^{(2)})
 h_a(x_3,x_2(1-r^2),b_3,b_2)S_t(x_2)rr_V\right.\nonumber\\
 &&\left.\times\left(\epsilon^{n\bar n\epsilon_D^{*T}\epsilon_V^{*T}}((1+x_2)\phi_V^a(x_3)
 +(1-x_2)\phi_V^v(x_3))\right.\right.\nonumber\\
 &&\left.\left.-i\epsilon_D^{*T}
 \cdot\epsilon_V^{*T}((1-x_2)\phi_V^a(x_3)+(1+x_2)\phi_V^v(x_3))\right)
\right].
 \label{texc}
 \end{eqnarray}
The evolution factors in the amplitudes are the same as those
in~(\ref{hd1}) and~(\ref{hd2}) after substituting $S_V(t)$ for
$S_P(t)$. For the nonfactorizable amplitudes, the factorization
formulas involve the kinematic variables of all the three mesons.
Their expressions are
\begin{eqnarray}
 {\cal M}_{\rm ext}^T&=& 16\pi\sqrt{2N_c} C_Fm_B^4 \int_0^1
 [dx]\int_0^{1/\Lambda} b_1 db_1 b_3 db_3
 \phi_B(x_1,b_1)\phi_{D}^T(x_2)
 \nonumber \\
 & &\times r_V\left[E_b(t_b^{(1)})h^{(1)}_b(x_i,b_i)\right.\nonumber\\
 &&\left.\times\left(\epsilon^{n\bar n\epsilon_D^{*T}\epsilon_V^{*T}}x_3(\phi_V^a(x_3)-\phi_V^v(x_3))+i\epsilon_D^{*T}
 \cdot\epsilon_V^{*T}x_3(\phi_V^a(x_3)-\phi_V^v(x_3))\right)\right.\nonumber\\
 &&\left.+E_b(t_b^{(2)})h^{(2)}_b(x_i,b_i)\right.\nonumber\\
 && \times\left\{\epsilon^{n\bar
 n\epsilon_D^{*T}\epsilon_V^{*T}}((1-x_3)(1-2r)\phi_V^a(x_3)+(x_3-1)\phi_V^v(x_3))\right.\nonumber\\
 && \left.  -i\epsilon_D^{*T}
 \cdot\epsilon_V^{*T}((x_3-1)\phi_V^a(x_3)+(1-2r)(1-x_3)\phi_V^v(x_3))\right\} \bigg]\;,
 \label{mext}\\
 {\cal M}_{\rm int}^T&=& 16\pi\sqrt{2N_c} C_Fm_B^4 \int_0^1
 [dx]\int_0^{1/\Lambda}b_1 db_1 b_2 db_2
 \phi_B(x_1,b_1)\phi_D^T(x_2)
 \nonumber \\
 & &\times r\left[E_d(t_d^{(2)})h^{(2)}_d(x_i,b_i)\right.\nonumber\\
 &&\left.\times\left(\epsilon^{n\bar n\epsilon_D^{*T}\epsilon_V^{*T}}((x_2-1)\phi_V^t(x_3)
 +r_V(\phi_V^a(x_3)+\phi_V^v(x_3)))\right.\right.\nonumber\\
 &&\left.\left.-i\epsilon_D^{*T}\cdot\epsilon_V^{*T}((1-x_2)\phi_V^t(x_3)+r_V(\phi_V^a(x_3)
 +\phi_V^v(x_3)))\right)\right.\nonumber\\
 &&\left.+E_d(t_d^{(1)})h^{(1)}_d(x_i,b_i)\right.\nonumber\\
 &&\left.\times\left(\epsilon^{n\bar
 n\epsilon_D^{*T}\epsilon_V^{*T}}(2r_V(x_3-1)\phi_V^a(x_3)-x_2\phi_V^t(x_3))\right.\right.\nonumber\\
 &&\left.\left.-i\epsilon_D^{*T}
 \cdot\epsilon_V^{*T}(x_2\phi_V^t(x_3)+2r_V(x_3-1)\phi_V^v(x_3))\right)\right]\;,
 \label{mint}\\
 {\cal M}_{\rm exc}^T&=& 16 \pi\sqrt{2N_c} C_Fm_B^4 \int_0^1
 [dx]\int_0^{1/\Lambda}b_1 db_1 b_2 db_2
 \phi_B(x_1,b_1)\phi_{D}^T(x_2)
 \nonumber \\
 & &\times \left\{E_f(t_f^{(1)})h^{(1)}_f(x_i,b_i)\phi_V^t(x_3)\right.\nonumber\\
 &&\left.\times\left(\epsilon^{n\bar n\epsilon_D^{*T}\epsilon_V^{*T}}(x_2r^2-r_V^2x_3)-i\epsilon_D^{*T}
 \cdot\epsilon_V^{*T}(x_2r^2+r_V^2x_3)\right)\right.\nonumber\\
 &&\left.+E_f(t_f^{(2)})h^{(2)}_f(x_i,b_i)\right.\nonumber\\
 && \times\left[\epsilon^{n\bar
 n\epsilon_D^{*T}\epsilon_V^{*T}}(-2rr_V\phi_V^a(x_3)-r^2(x_2-1)\phi_V^t(x_3)+r_V^2(x_3-1)\phi_V^t(x_3))\right.
 \nonumber\\
 &&\left. +i\epsilon_D^{*T}
 \cdot\epsilon_V^{*T}(((x_2-1)r^2+r_V^2(x_2-1))\phi_V^t(x_3)+2rr_V\phi_V^v(x_3))\right]\bigg\}\;. \label{mexc}
\end{eqnarray}
The $h's$ and $h^{(j)}$ functions in the amplitudes here are the
same as defined in (\ref{hd1},\ref{hd2},\ref{hb}-\ref{hf}).

If carrying
 out the substitutions in~(\ref{D*replace}) and (\ref{Vreplace})
 together, we can get the final decay amplitude of each polarization for $B\to D^*V$ decays.

 \section{Numerical Calculations and discussions}\label{section:results}

  In this section the numerical results of our calculation will be
  given.
 The parameters of $D_{(s)}^{(*)}$ meson we use are
 \begin{eqnarray}
m_{D^-}&=&1.869\mbox{GeV}, \qquad m_{D_s^-} = 1.968\mbox{GeV},  \nonumber \\
m_{D^{*-}} &=& 2.010\mbox{GeV}, \qquad m_{D_s^{*-}} = 2.112\mbox{GeV},  \nonumber \\
f_{D^-} &=& 223 \mbox{MeV},    \qquad f_{D_s^-} = 274 \mbox{MeV}.
\end{eqnarray}
A lot of study has been made on the decay constants of $D_{(s)}$
mesons. Here we take the values from ref.~\cite{D meson decay
constant CLEO }. Since there are no experimental results of the
decay constants of $D_{(s)}^*$ mesons, we use  the relations between
$f_D$ and $f_{D^*}$ derived from HQET:
\begin{eqnarray}
f_{D^{*-}} = \sqrt{\frac{m_{D^-}}{m_{D^*-}}}f_{D^-}, \qquad
f_{D_s^{*-}} = \sqrt{\frac{m_{D_s^-}}{m_{D_s^*-}}}f_{D_s^-},
\end{eqnarray}
which is different from ref.\cite{Weak decays of J/psi}.

 With the $D$ meson wave functions we get, we can calculate the
 amplitudes easily. And the decay width can be got by
 \begin{equation}
 \Gamma=\frac{1}{32\pi}G_F^2|V_{cb}|^2|V_{ud}|^2m_B^7(1-r^2)|A|^2\;
 ,
 \end{equation}
 with $A$ the decay amplitude defined in eqs.(\ref{B_to_Dpi-}-\ref{Bs_to_etas})
 The branch ratio is
 \begin{equation}
 Br=\Gamma\hbar/\tau_{B_{(s)}},
 \end{equation}
 with $\tau_{B_{(s)}}$ as the life time of $B_{(s)}$ meson. We take
 $\tau_{B^-}=1.674\times10^{-12}s$, $\tau_{\bar B^0}=1.542\times10^{-12}s$,
  $\tau_{\bar B^0_s}=1.466\times10^{-12}s$, and
  $G_F=1.16639\times10^{-5}$.

 \subsection{results of fitting}

 \begin{table}
 \caption{The smallest $\chi^2$ for each kind of the D meson DAs }
 \label{fitresults}
 \begin{center}
 \begin{tabular}{l  cccccc}
 \hline\hline
 \ \ \           &$\Phi^{(Gen)}$ &$\Phi^{(MGen)}$ &$\Phi^{(KLS)}$  &$\Phi^{(GN)}$  &$\Phi^{(KKQT)}$  &$\Phi^{(Huang)}$    \\
 \hline
 \ \ \ $\chi^2_{min}$ &$34.1$  &$33.6$   &$156.9$   &$112.3$ &$45.0$     &$41.9$     \\
 \hline\hline
 \end{tabular}
 \end{center}
 \end{table}

\begin{figure}
 \includegraphics[scale=1]{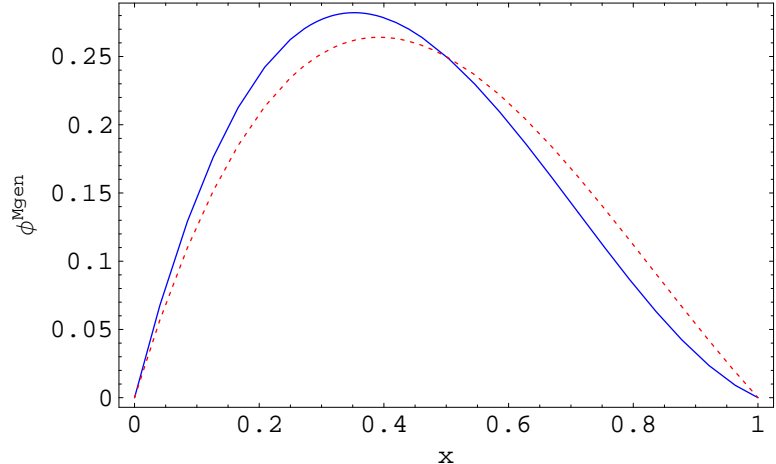}
 \caption{The D meson distribution amplitude $\phi_{D}^{MGen}(0.8,0)$  ( the blue, solid line)
 and
  $D_s$ meson distribution amplitude $\phi_{D_s}^{MGen}(0.5,0)$ (the red, dotted line)}
 \label{fig:D_wavefunction}
 \end{figure}

 Since the $B \to DP$ decay channels have been measured in high
 precision, we use these experimental results to fit out the
 parameters of the candidate D meson DAs. Here we don't use the
  experimental results containing $\eta$ or $\eta'$ in the final states because there are
   uncertainties from the mixing. The six decay channels
 we used to fit out the D meson wave function parameters are $B^-\to D^0 \pi^-$, $B^- \to D^0
 K^-$,$\bar{B^0} \to D^+ \pi^-$, $\bar{B^0}\to D^+ K^-$, $\bar{B^0}\to D^0 \pi^0$,
 $\bar{B^0} \to D^0 \bar{K^0}$. The experimental results of these channels
 are listed in table~\ref{PPresults}, which are from ref.\cite{experiment:B_to_D}.
 The formula we used for fitting is
 \begin{eqnarray}
 \chi^2=\sum_i\frac{(Br^{ex}_i-Br^{th}_i)^2}{\sigma_i^2}.
 \end{eqnarray}
 The $i$ means the summation over the six decay channels. $Br^{ex}_i$($Br^{th}_i$) is the
 experimental (theoretical) value of branch ratio, and $\sigma_i$
 is the uncertainty of the experimental value.
 In table~\ref{fitresults} we list the smallest $\chi^2$ we get
 for all the D meson DAs. Easy to see, except $\Phi^{(KLS)}$ and
 $\Phi^{(GN)}$, all the other DAs have a small $\chi^2_{min}$. The
 $\Phi^{(MGen)}$ is the best one, with its
 parameters fixed as $C_D=0.8$, $\omega=0.1$. We will use this D
 meson wave function for the following numerical calculations of all
 the decay channels.
 For the $D_s$ meson, we
 use $C_D=0.5$, $\omega=0.2$, with a little SU(3) breaking effect. In this case, we can see
  from
 Fig.~\ref{fig:D_wavefunction} that the $\bar s$ quark in $D_s$ meson has a little larger
 momentum fraction than the $\bar d$ quark in the $D$ meson, which characterize the little
 larger mass of $s$ quark.

 Because the mass difference between the vector meson
 $D^*_{(s)}$ and pseudoscalar meson $D_{(s)}$ is small, so we adopt the same DA for
 them.

 \subsection{Results for all the related channels and discussions}

 \begin{table}[t]
\caption{Branching ratios of $B_{(s)}\to DP$ decays  calculated in
pQCD approach with experimental data (in units of $10^{-4}$) }
 \label{PPresults}
\begin{center}
\begin{tabular}{l  cc} \hline \hline
 & Experimental results   &  Our results    \\
\hline
$B^- \to D^0 \pi^-$          &$47.5\pm 1.9 $                    &$51.1_{-20.7-7.5-1.5}^{+29.5+4.3+1.5}$   \\
$B^- \to D^0 K^-$            &$3.83\pm 0.45 $                    &$4.00_{-1.64-0.93-0.12}^{+2.35+0.63+0.12}$   \\
$\bar{B}^0 \to D^+ \pi^-$    &$26.5\pm 1.5 $                    &$26.9_{-11.7-7.3-0.8}^{+17.8+5.5+0.8}$   \\
$\bar{B}^0 \to  D^+ K^-$     &$2.04\pm 0.57$                     &$2.43_{-1.01-0.71-0.07}^{+1.56+0.63+0.07}$   \\
$\bar{B}^0 \to D^0 \pi^0$    &$2.61\pm 0.25 $                    &$1.98_{-0.66-0.63-0.06}^{+0.67+0.51+0.06}$   \\
$\bar{B}^0\to D^0 \bar{K}^0$ &$0.523 \pm 0.066 $       &$0.22_{-0.07-0.06-0.01}^{+0.08+0.06+0.01}$   \\
$\bar{B}^0 \to D^0 \eta$     &$2.02\pm 0.21 $                    &$2.46_{-0.77-0.30-0.07}^{+0.97+0.30+0.07}$   \\
$\bar{B}^0 \to D^0 \eta'$    &$1.26\pm 0.21 $                    &$1.65_{-0.52-0.20-0.05}^{+0.65+0.20+0.05}$   \\
$\bar{B}^0 \to D_s^+ K^-$    &$0.269\pm 0.054  $       &$0.48_{-0.12-0.11-0.01}^{+0.16+0.11+0.01}$   \\
\hline
$\bar{B_s}^0 \to D^+ \pi^-$      &                          &$2.32_{-0.61-0.39-0.07}^{+0.82+0.32+0.07}\times10^{-2}$ \\
$\bar{B_s}^0 \to D^0 \pi^0$      &                          &$1.15_{-0.29-0.20-0.04}^{+0.36+0.19+0.04}\times10^{-2}$ \\
$\bar{B_s}^0 \to D^0 \bar K^0$   &                          &$3.96_{-1.25-0.99-0.12}^{+1.49+0.88+0.11}$  \\
$\bar{B_s}^0 \to D^0 \eta$       &                          &$0.14_{-0.05-0.03-0.00}^{+0.05+0.03+0.00}$ \\
$\bar{B_s}^0 \to D^0 \eta'$      &                          &$0.32_{-0.10-0.04-0.01}^{+0.11+0.03+0.01}$ \\
$\bar{B_s}^0 \to D_s^+ \pi^-$    & $38\pm3 \pm 13$          &$21.3_{-8.1-6.8-0.6}^{+11.4+6.9+0.6}$ \\
$\bar{B_s}^0 \to D_s^+ K^-$      &                          &$1.71_{-0.65-0.55-0.05}^{+0.92+0.58+0.05}$\\
 \hline \hline
\end{tabular}
\end{center}
\end{table}

 Our numerical results are listed in table~\ref{PPresults},
 table~\ref{PVresults}, table~\ref{VPresults} and table~\ref{VVresults}.
 The first error in these entries are caused by the hadronic parameters in
 $B_{(s)}$ meson wave function (the decay constant and the shape
 parameter). We take $f_B=0.19\pm 0.025$, $\omega_b=0.40\pm0.05$
 and $f_{B_s}=0.24\pm 0.03$, $\omega_b^{B_s}=0.50\pm0.05$. The second error
 arise from the higher order perturbative QCD corrections: the choice  of the hard scales, defined in
 (\ref{hardscale1}) and (\ref{hardscale2}), which vary from $0.75t$ to $1.25t$,
  and the uncertainty of $\Lambda_{QCD}^{(4)}=0.25\pm0.05$. The third error is from the uncertainties
 of the CKM matrix elements.
 In our calculation, we use
 \begin{equation}
 V_{cb}=\left(41.61_{-0.63}^{+0.62}\right)\times10^{-3}\;,\;V_{ud}=0.97385_{-0.00023}^{+0.00024}\;,
 \;V_{us}=0.22715_{-0.00100}^{+0.00101}\;.
 \end{equation}
Among  them, the hadronic inputs always gives rise
 to the largest uncertainty, and the CKM matrix elements contribute little.

 \begin{table}[t]
\caption{Branching ratios of $B_{(s)}\to DV$ decays  calculated in
pQCD approach with experimental data   (in units of $10^{-4}$) }
 \label{PVresults}
\begin{center}
\begin{tabular}{l  cc} \hline \hline
 & Experimental results   &  Our results    \\
\hline
$B^- \to D^0 \rho^-$                    &$134\pm18$                  &$113_{-45.9-26.1-3.3}^{+68.1+18.8+3.3}$   \\
$B^- \to D^0 K^{*-}$                    &$5.29\pm0.45$               &$6.49_{-2.68-1.58-0.20}^{+3.86+0.12+0.20}$   \\
$\bar{B}^0 \to D^+ \rho^-$              &$75\pm12$                   &$69.6_{-29.7-17.8-2.0}^{+45.0+13.1+2.0}$   \\
$\bar{B}^0 \to  D^+ K^{*-}$             &$4.60\pm0.78$               &$4.07_{-1.69-1.11-0.12}^{+2.61+0.94+0.12}$   \\
$\bar{B}^0 \to D^0 \rho^0$              &$2.91\pm0.40$               &$1.79_{-0.62-0.59-0.05}^{+0.65+0.46+0.05}$   \\
$\bar{B}^0 \to D^0 \omega$              &$2.60\pm0.29$               &$4.25_{-1.21-0.73-0.12}^{+1.58+0.85+0.12}$   \\
$\bar{B}^0\to D^0 \bar{K}^{*0}$         &$0.423\pm0.064 $  &$0.26_{-0.08-0.07-0.01}^{+0.10+0.08+0.01}$    \\
$\bar{B}^0 \to D_s^+ K^{*-}$            &$<8$                        &$1.94_{-0.47-0.38-0.06}^{+0.66+0.41+0.06}$   \\
\hline
$\bar{B_s}^0 \to D^+ \rho^-$            &                          &$0.11_{-0.03-0.02-0.00}^{+0.03+0.02+0.00}$ \\
$\bar{B_s}^0 \to D^0 \rho^0$            &                          &$\left(5.42_{-1.39-1.11-0.16}^{+1.57+0.96+0.16}\right)\times10^{-2}$ \\
$\bar{B_s}^0 \to D^0  \bar K^{*0}$      &                          &$4.52_{-1.43-1.30-0.13}^{+1.70+1.15+0.13}$  \\
$\bar{B_s}^0 \to D^0 \omega$            &                          &$\left(4.71_{-1.18-0.92-0.14}^{+1.35+0.88+0.14}\right)\times10^{-2}$ \\
$\bar{B_s}^0 \to D^0 \phi$              &                          &$0.30_{-0.10-0.08-0.01}^{+0.11+0.07+0.01}$ \\
$\bar{B_s}^0 \to D_s^+ \rho^-$          &                          &$51.0_{-19.2-16.2-1.48}^{+26.9+16.6+1.48}$ \\
$\bar{B_s}^0 \to D_s^+ K^{*-}$          &                          &$3.02_{-1.16-0.91-0.10}^{+1.62+0.88+0.10}$\\
 \hline \hline
\end{tabular}
\end{center}
\end{table}

 \begin{table}[t]
\caption{Branching ratios of $B_{(s)}\to D^*P$ decays  calculated in
pQCD approach with experimental data (in units of $10^{-4}$) }
 \label{VPresults}
\begin{center}
\begin{tabular}{l  c c} \hline \hline
 & Experimental results   &  Our results    \\
\hline
$B^- \to D^{*0} \pi^-$          &$52.8\pm2.8$                   &$50.4_{-20.4-7.3-1.5}^{+29.4+4.4+1.5}$   \\
$B^- \to D^{*0} K^-$            &$3.6\pm1.0$                    &$3.98_{-1.62-0.92-0.12}^{+2.33+0.62+0.12}$   \\
$\bar{B}^0 \to D^{*+} \pi^-$    &$26.2\pm1.3$                   &$26.0_{-11.4-7.09-0.76}^{+17.3+5.34+0.75}$   \\
$\bar{B}^0 \to  D^{*+} K^-$     &$2.04\pm0.47$                  &$2.37_{-0.99-0.69-0.07}^{+1.52+0.62+0.07}$   \\
$\bar{B}^0 \to D^{*0} \pi^0$    &$1.71\pm0.28$                  &$2.11_{-0.69-0.62-0.06}^{+0.76+0.52+0.06}$   \\
$\bar{B}^0 \to D^{*0} \eta$     &$1.80\pm0.31$                  &$2.60_{-0.83-0.33-0.08}^{+1.01+0.29+0.08}$   \\
$\bar{B}^0 \to D^{*0} \eta'$    &$1.21\pm0.40$                  &$1.74_{-0.55-0.22-0.05}^{+0.68+0.19+0.05}$   \\
$\bar{B}^0\to D^{*0} \bar{K}^0$ &$0.36\pm0.12$                  &$0.24_{-0.07-0.06-0.01}^{+0.09+0.06+0.01}$    \\
$\bar{B}^0 \to D_s^{*+} K^-$    &$0.200\pm0.064 $     &$0.47_{-0.12-0.11-0.01}^{+0.16+0.12+0.01}$   \\
\hline
$\bar{B_s}^0 \to D^{*+} \pi^-$    &        &$\left(2.27_{-0.60-0.38-0.01}^{+0.81+0.31+0.01}\right)\times10^{-2}$ \\
$\bar{B_s}^0 \to D^{*0} \pi^0$    &        &$\left(1.13_{-0.28-0.19-0.03}^{+0.35+0.18+0.03}\right)\times10^{-2}$ \\
$\bar{B_s}^0 \to D^{*0} \bar K^0$ &        &$4.27_{-1.35-1.05-0.12}^{+1.54+0.88+0.12}$ \\
$\bar{B_s}^0 \to D^{*0} \eta$     &        &$0.15_{-0.05-0.03-0.00}^{+0.06+0.03+0.00}$ \\
$\bar{B_s}^0 \to D^{*0} \eta'$    &        &$0.33_{-0.10-0.03-0.01}^{+0.12+0.03+0.01}$ \\
$\bar{B_s}^0 \to D_s^{*+} \pi^-$  &        &$24.2_{-7.2-7.7-0.7}^{+11.2+7.8+0.7}$ \\
$\bar{B_s}^0 \to D_s^{*+} K^-$    &        &$1.65_{-0.63-0.53-0.05}^{+0.90+0.56+0.05}$ \\
 \hline \hline
\end{tabular}
\end{center}
\end{table}

The first six channels in table~\ref{PPresults} are input values of
the $\chi^2$ fit program. Although we get a reasonable $\chi^2$ in
the fit,
   the branching ratio of $\bar B^0\to D^0\bar K^0$ is about only half
 of the experimental value.  Comparing with the color suppressed diagrams,
 the annihilation diagrams contributes little for $\bar B^0\to D^0\bar K^0$ and $\bar B^0\to D^0\pi^0$.
 So we can take them for a compare. They have different CKM elements ($V_{cb}V^*_{us}$ for the former
 and $V_{cb}V^*_{ud}$ for the latter). Taking the factor $\frac{1}{\sqrt{2}}$ in flavor
 wave function of $\pi$ meson into account, the Br($\bar B^0\to D^0\bar K^0$) is
 roughly one tenth of Br($\bar B^0\to D^0\pi^0$). So this value of Br($\bar B^0\to D^{0}
 \bar K^0$) is theoretically reasonable. The similar argument is valid for Br($\bar
 B^0\to D^{*0}\bar K^0$).

Although we use only six $B\to DP$ channels to fix the D meson wave
function, the results of most other channels especially those $B\to
DV$ and $D^*P$ channels agree very well with the current
experimental measurements.
   Easy to find that, Br($B\to D\omega$) is twice larger than Br($B\to D\rho$),
 while their experimental results are near to each other.
 Both these two channels receive contributions from the
 color suppressed diagrams and annihilation diagrams. They are at the same order
 magnitude for these two processes. For color suppressed diagrams, the
 $d\bar d$ of the flavor part contributes. While the $u\bar u$
 part contributes to the annihilation diagrams. Amplitudes of
 these two kinds of diagrams have the same sign
   in the  $B\to D\omega$ decay but different sign in the $B\to
   D\rho$ decay due to isospin.
 Similar situation exists for Br($B\to D^*\omega$) and Br($B\to D^*\rho$).

The $\bar B^0 \to D_s^+K^-$ decay is a kind of pure annihilation
type decays dominant by W exchange diagram. Our result is larger
than the experiments and also larger than the previous pQCD
calculations \cite{anni} due to the change of D meson wave
functions. The annihilation type diagrams are power suppressed in
pQCD approach, which is more sensitive to the hadronic wave
functions.

 \begin{table}[t]
\caption{Predicted branching ratios of $B_{(s)}\to D^*V$ decays with
experimental data (in units of $10^{-4}$) together with the
percentage of transverse polarizations $R_T$}
 \label{VVresults}
\begin{center}
\begin{tabular}{lcccc} \hline \hline

 & Experimental BRs &   &  BRs in pQCD &  $R_T$  \\
\hline
$B^- \to D^{*0} \rho^-$                    &$$ &$$                       &$117_{-48.0-27.0-3.4}^{+71.0+19.9+3.4}$ &$0.04$   \\
$B^- \to D^{*0} K^{*-}$                    &$8.3\pm1.5$ &$0.14$          &$6.82_{-2.80-1.65-0.21}^{+4.14+1.22+0.21}$ &$0.06$   \\
$\bar{B}^0 \to D^{*+} \rho^-$              &$$ &$$                       &$79.4_{-34.5-21.2-2.3}^{+52.6+15.9+2.3}$ &$0.15$   \\
$\bar{B}^0 \to  D^{*+} K^{*-}$             &$3.20\pm0.67$ &$$            &$4.88_{-2.08-1.41-0.15}^{+3.18+1.16+0.15}$ &$0.19$   \\
$\bar{B}^0 \to D^{*0} \rho^0$              &$3.73\pm0.99$ &$$           &$3.69_{-1.23-0.63-0.11}^{+1.52+0.57+0.11}$ &$0.47$   \\
$\bar{B}^0 \to D^{*0} \omega$              &$2.68\pm0.50$ &$$            &$5.83_{-1.84-1.10-0.17}^{+2.14+0.84+0.17}$ &$0.24$   \\
$\bar{B}^0\to D^{*0} \bar K^{*0}$          &$<0.69$ &$$                  &$0.51_{-0.17-0.10-0.02}^{+0.20+0.08+0.02}$ &$0.44$    \\
$\bar{B}^0 \to D_s^{*+} K^{*-}$            &$$ &$$                       &$1.97_{-0.49-0.43-0.06}^{+0.58+0.37+0.06}$ &$0.02$   \\
\hline
$\bar{B_s}^0 \to D^{*+} \rho^-$            &&            &$0.11_{-0.03-0.02-0.00}^{+0.03+0.02+0.00}$ &$0.01$ \\
$\bar{B_s}^0 \to D^{*0} \rho^0$            &&            &$\left(5.35_{-1.39-1.05-0.16}^{+1.44+0.96+0.16}\right)\times 10^{-2}$ &$0.01$ \\
$\bar{B_s}^0 \to D^{*0}  \bar K^{*0} $     &&            &$8.43_{-2.67-1.83-0.25}^{+3.30+1.68+0.24}$ &$0.43$  \\
$\bar{B_s}^0 \to D^{*0} \omega$            &&            &$\left(4.60_{-1.11-0.84-0.14}^{+1.50+0.89+0.14}\right)\times 10^{-2}$ &$0.01$ \\
$\bar{B_s}^0 \to D^{*0} \phi$              &&            &$0.51_{-0.17-0.11-0.02}^{+0.19+0.09+0.02}$ &$0.37$ \\
$\bar{B_s}^0 \to D_s^{*+} \rho^-$          &&            &$56.9_{-21.6-18.3-1.66}^{+30.4+19.1+1.65}$ &$0.13$ \\
$\bar{B_s}^0 \to D_s^{*+} K^{*-}$          &&            &$3.47_{-1.35-1.06-0.11}^{+1.96+1.07+0.11}$ &$0.17$\\
 \hline \hline
\end{tabular}
\end{center}
\end{table}

For the decays $B\to D^*_{(s)}V$ in table~\ref{VVresults}, we also
estimate the ratios
 of transverse polarized contribution $R_T=|A_T|^2/(|A_T|^2+|A_L|^2)$.
  We should mention that,
these results are just indicative, because transverse polarizations
are power suppressed by $r_V$ or $r$ to make it more sensitive to
small parameters and higher order contributions than
   the longitudinal contribution.
   Although the transverse polarization is suppressed in $B\to D^*V$
 decays, in some channels, such as $B\to D^{*0}\rho^0$ and $B\to D^{*0}\bar K^{*0}$,
 etc, it has 40\% contributions. The reason is that the dominant
 contribution in these channels is from ${\cal M}_{int}$ in eq.(\ref{md}), which is
 $x_3$ suppressed, while the transverse contribution in
 eq.(\ref{mint}) is only $r$ suppressed. They are comparable
 numerically to make a large contribution for transverse
 polarizations in these color suppressed channels. This mechanism is
 different from those charmless $B$ decays where the dominant
 transverse polarizations are from the space like penguin (penguin
 annihilation)
 contributions \cite{trans}. Here the annihilation type
 contributions are mainly from W exchange diagrams contributing
 little to transverse polarizations.

For those previous calculated channels in pQCD approach \cite{PQCD B
to D,Kurimoto:2002sb}, our results are slightly different due to
parameter changes. Most of the $B^0(B^\pm)$ decay channels are
measured by the two B
 factories, which are consistent with our calculations. For the
 $B_s$ decays, only one channel is measured. Our predictions will
 soon be tested by the LHCb experiments.

For comparison with other methods, we also give the form factors  at
the maximal recoil
 \begin{equation}
 \xi^{B\to D}_+=0.52_{-0.12-0.07}^{+0.15+0.05}\;,\;
 \xi^{B_s\to D_s}_+=0.46_{-0.09-0.08}^{+0.11+0.07}.
 \end{equation}
These are comparable with other methods \cite{BtoDform-factor}.

 If applying the naive factorization approach, we can get
 \begin{eqnarray}
 A(\bar B^0\to
 D^+\pi^-)&=&i\frac{G_F}{\sqrt{2}}V_{cb}V^*_{ud}(M_B^2-M_D^2)f_{\pi}F^{B\to
 D}(M_{\pi}^2)a_1(D\pi),\label{a1}\\
 \sqrt{2}A(\bar B^0\to
 D^0\pi^0)&=&-i\frac{G_F}{\sqrt{2}}V_{cb}V^*_{ud}(M_B^2-M_{\pi}^2)f_DF^{B\to
 \pi}(M_D^2)a_2(D\pi).\label{a2}
 \end{eqnarray}
 Substituting our results for $A(\bar B^0\to D^+\pi^-)$ and $A(\bar B^0\to
 D^0\pi^0)$ in eq.(\ref{a1}) and (\ref{a2}), we can extract the BSW
 parameters $a_1$ and $a_2$ from our pQCD approach
 \begin{equation}
 |a_2/a_1|=0.49\;,\;Arg(a_2/a_1)=-37.6^{\circ}.
 \end{equation}
 If the annihilation diagrams' contribution is excluded, the
 results are
 \begin{equation}
 |a_2/a_1|=0.54\;,\;Arg(a_2/a_1)=-59.0^{\circ}.
 \end{equation}
 Indeed, the large $|a_2/a_1|$ implies that the color suppressed
 decays not very suppressed as previous expected \cite{comment on
 color-suppressed}. The relative strong phase between the two
 contributions is not small as naive expectations.
  These results are
 consistent with recent direct studies from experiments
 \cite{jiang}. But the difference is that our results come from
 direct dynamical calculation not from fit.

 \section{Conclusion}\label{section:conclusion}

 In this paper, we calculate the branch ratios of $B_{(s)} \to D_{(s)}P$,
$D_{(s)}^*P$, $D_{(s)}V$ and $D_{(s)}^*V$ channels, with the D meson
wave function obtained
 through fitting. We also calculate the ratios of transverse
 polarized contributions in $B\to D^*V$ decays. Most of the results
 agree well with the experiments. It seems that there's a
 disagreement with the experimental data in the relative size  of branching ratios
 for $B\to
 D^{(*)}\rho$ and $B\to D^{(*)}\omega$. Some channels of the $B\to D^*V$
 decays may receive large contribution from the transverse
 polarization. The results of $\bar B_s^0\to D_{(s)} P$, $D_{(s)}  V$,
 $D_{(s)}^{*} P $, $D_{(s)}^{*} V$ decays
 will be
 tested in the future experiments.

 \section*{Acknowledgments}
 We thank Wei Wang and Yu-Ming Wang a lot, for the  fruitful
 discussions. This work is
partly supported by National  Science Foundation of China under the
Grant Numbers 10735080, 10625525 and 10525523.

 \appendix
 \section{pQCD functions}

 Jet function appears in (\ref{ext})-(\ref{exc}),
 (\ref{text})-(\ref{texc}) is
 \begin{equation}
 S_t(x)=\frac{2^{1+2c}\;\Gamma(3/2+c)}{\sqrt{\pi}\;\Gamma(1+c)}[x(1-x)]^c
 \end{equation}
 the value of c in the above equation is $0.5$ in this paper. And
 the $S_j(x_i)$($j=B,C,P$ or $V$) functions in Sudakov form factors in
 (\ref{Sudakov1}) and (\ref{Sudakov2}) are
 \begin{eqnarray}
 S_B(t)&=&s\left(x_1\frac{m_{B}}{\sqrt{2}},b_1\right)+2\int^t_{1/b_1}\frac{d\bar \mu}{\bar
 \mu}\gamma_q(\alpha_s(\bar \mu)),\\
 S_C(t)&=&s\left(x_2\frac{m_{B}}{\sqrt{2}},b_2\right)+2\int^t_{1/b_2}\frac{d\bar \mu}{\bar
 \mu}\gamma_q(\alpha_s(\bar \mu)),\\
 S_V(t)&=&S_P(t)=s\left(x_3\frac{m_{B}}{\sqrt
2},b_3\right)+s\left((1-x_3)\frac{m_{B}}{\sqrt
2},b_3\right)+2\int^t_{1/b_3}\frac{d\bar \mu}{\bar
\mu}\gamma_q(\alpha_s(\bar \mu)),
 \end{eqnarray}
 with the quark anomalous dimension $\gamma_q=-\alpha_s/\pi$. The
explicit form for the  function $s(Q,b)$ is:
\begin{eqnarray}
s(Q,b)&=&~~\frac{A^{(1)}}{2\beta_{1}}\hat{q}\ln\left(\frac{\hat{q}}
{\hat{b}}\right)-
\frac{A^{(1)}}{2\beta_{1}}\left(\hat{q}-\hat{b}\right)+
\frac{A^{(2)}}{4\beta_{1}^{2}}\left(\frac{\hat{q}}{\hat{b}}-1\right)
-\left[\frac{A^{(2)}}{4\beta_{1}^{2}}-\frac{A^{(1)}}{4\beta_{1}}
\ln\left(\frac{e^{2\gamma_E-1}}{2}\right)\right]
\ln\left(\frac{\hat{q}}{\hat{b}}\right)
\nonumber \\
&&+\frac{A^{(1)}\beta_{2}}{4\beta_{1}^{3}}\hat{q}\left[
\frac{\ln(2\hat{q})+1}{\hat{q}}-\frac{\ln(2\hat{b})+1}{\hat{b}}\right]
+\frac{A^{(1)}\beta_{2}}{8\beta_{1}^{3}}\left[
\ln^{2}(2\hat{q})-\ln^{2}(2\hat{b})\right],
\end{eqnarray} where the variables are defined by
\begin{eqnarray}
\hat q\equiv \mbox{ln}[Q/(\sqrt 2\Lambda)],~~~ \hat b\equiv
\mbox{ln}[1/(b\Lambda)], \end{eqnarray} and the coefficients
$A^{(i)}$ and $\beta_i$ are \begin{eqnarray}
\beta_1=\frac{33-2n_f}{12},~~\beta_2=\frac{153-19n_f}{24},\nonumber\\
A^{(1)}=\frac{4}{3},~~A^{(2)}=\frac{67}{9}
-\frac{\pi^2}{3}-\frac{10}{27}n_f+\frac{8}{3}\beta_1\mbox{ln}(\frac{1}{2}e^{\gamma_E}),
\end{eqnarray}
$n_f$ is the number of the quark flavors and $\gamma_E$ is the Euler
constant. We will use the one-loop running coupling constant, i.e.
we pick up the four terms in the first line of the expression for
the function $s(Q,b)$.

\end{document}